\documentclass[aps,prd,10pt,nofootinbib,twocolumn,eqsecnum,showpacs,showkeys,superscriptaddress,preprintnumbers,altaffilletter]{revtex4-2}
\usepackage{graphicx}
\usepackage{amsmath}
\usepackage{multirow}
\usepackage{hyperref}
\usepackage{soul}
\usepackage{graphicx}
\usepackage{dcolumn}
\usepackage{amssymb}
\usepackage{amsfonts}
\usepackage{amsbsy}
\usepackage{color}
\usepackage{rotating}
\usepackage[english]{babel}
\usepackage{multirow}
\usepackage{float}
\usepackage{cleveref}
\usepackage{aas_macros}
\usepackage{diagbox}
\usepackage{leftidx}
\usepackage{float}
\newcommand{\be}{\begin{equation}}
\newcommand{\ee}{\end{equation}}
\newcommand{\bea}{\begin{eqnarray}}
\newcommand{\eea}{\end{eqnarray}}
\newcommand{\beal}{\begin{aligned}}
\newcommand{\eeal}{\end{aligned}}
\newcommand{\bi}{\begin{itemize}}
\newcommand{\ei}{\end{itemize}}

\hypersetup{
    colorlinks=true,
    linkcolor=blue,
    filecolor=magenta,
    urlcolor=cyan,
    citecolor=cyan
}

\begin{document}

\title{DHOST gravity in Ultra-diffuse galaxies - Part I: the case of NGC1052-DF2}
%\title{Lacking dark matter Ultra-diffuse galaxies as test arena for DHOST gravity}

\author{E. Laudato}
\email{enrico.laudato@phd.usz.edu.pl}
\affiliation{Institute of Physics, University of Szczecin, Wielkopolska 15, 70-451 Szczecin, Poland}
\author{V. Salzano}
\email{vincenzo.salzano@usz.edu.pl}
\affiliation{Institute of Physics, University of Szczecin, Wielkopolska 15, 70-451 Szczecin, Poland}

\date{\today}

\begin{abstract}

The Ultra-Diffuse galaxy NGC1052-DF2 has recently been under intense scrutiny because from its kinematics it has revealed to be ``extremely deficient'' in dark matter, if not lacking it at all. This claim has raised many questions and solutions regarding the relationship between baryons and dark matter in Ultra-Diffuse galaxies. But there seems to be a quite unanimous belief that, if such very low dark matter content is confirmed and extended to other similar galactic objects, it might be a deathblow to theories which modify and extend General Relativity. Deficient dark matter galaxies thus represent a fertile ground to test both standard dark matter and modified gravity theories. In this work, we consider a specific Degenerate Higher-Order Scalar Tensor model to study the velocity dispersion of ten compact globular clusters-like objects associated with NGC1052-DF2 to infer the dynamical mass of the galaxy. Due to the partial breaking of the corresponding screening mechanism, this model can possibly have large cosmological scale effects influencing the dynamics of smaller structures like galaxies. We consider two scenarios: one in which the model only describes dark energy; and one in which it additionally entirely substitutes dark matter. We find that the best model to explain data is the one in which we have General Relativity and only stellar contribution. But while in former scenario General Relativity is still statistically (Bayesian) favoured, in the latter one the alternative model is as much successful and effective as General Relativity in matching observations. Thus, we can conclude that even objects like NGC1052-DF2 are not in contrast, and are not obstacles, to the study and the definition of a reliable alternative to General Relativity.
\end{abstract}

\maketitle

\section{Introduction}

The quest for understanding the true nature of the dark components of our Universe is expected to enter a new exciting era of deep insights due to the unprecedented quality data which missions like the \textit{James Webb Space Telescope}, \textit{Euclid} and the Square Kilometer Array will provide us in the very next future. In the wait for such new data, the present observational state-of-the-art \cite{Abdalla:2022yfr} teaches us a lot about both Dark Matter (DM) and Dark Energy (DE), but yet not enough to boast unequivocal claims about them.

In this work we are going to focus more specifically on DM. DM has a long history, starting from the very first claim for its existence \cite{1937ApJ....86..217Z}, passing through collecting decisive evidences \cite{1970ApJ...159..379R}, and arriving to confirmation(s) by the most varied cosmological and astrophysical probes \cite{Planck:2018vyg,eBOSS:2020yzd,DES:2021wwk}. Although nowadays many dedicated experiments have been realized and are led to detect DM (see, e.g. \cite{Durnford:2021mzg,XENON:2019zpr,Felcini:2018osp,DEAP:2019yzn,Bernabei:2013xsa}), we do not have any proof of any theorized DM candidate particles \cite{Bertone:2004pz}. Moreover, we also know that most (if not all) the theoretical scenarios we have to explain its nature, origin and dynamics, have many problems \cite{Bull:2015stt}.

In order to shed some light onto it, we decided to follow a very specific and (until not long ago) less-conventional approach: the DM problem (as well as the DE one) might arise due to the assumption of General Relativity (GR) as the ultimate theory of gravity in our standard cosmological picture. Thus, a possible solution could be to extend GR into a more general gravity theory. This is the approach of the so-called Extended Theories of gravity (ETGs). The ways in which GR might be extended are almost uncountable \cite{Clifton:2011jh}, \textit{in theory}, but \textit{practically} a gravity modification can not be done arbitrarily. Indeed, each and every proposed ETG must reduce to GR at Solar System scales, where GR has been proven to work perfectly. Moreover, in order to preserve the Equivalence Principle, there must be a screening mechanism \cite{Brax_2013,Koyama:2015vza,Burrage:2017qrf}, that suppresses at small scales any large-scale gravity modifications.

In \cite{10.1093/mnras/stac180}, we started to investigate a specific class \cite{PhysRevD.97.021301,Dima:2017pwp} of ETGs called Degenerate Higher-Order Scalar Tensor (DHOST) theories \cite{BenAchour:2016fzp,Langlois:2018dxi,Dima:2017pwp,Crisostomi:2016czh,PhysRevD.97.021301}, which represent a generalization of the Beyond Horndeski theories \cite{Gleyzes:2014qga} (which in turn generalize the well known Horndeski's theory of gravity \cite{Horndeski:1974wa}). The main characteristic of such a class which drove our attention, is that it can be characterized by a partial breaking of the corresponding screening mechanism, the so-called Vainshtein screening \cite{Vainshtein:1972sx}. This means that the gravity modifications which are introduced at large cosmological scales to specifically mimic GR-based DE models, might leak onto smaller astrophysical scales, and thus might play some role as ``effective'' DM. 

In \cite{10.1093/mnras/stac180} we tested the chosen model with a sample of sixteen high-mass galaxy clusters belonging to the Cluster Lensing and Supernova survey
with \textit{Hubble} (CLASH) program \cite{2012ApJS..199...25P} for which we had data available from two complementary probes, namely X-ray and strong-and-weak gravitational lensing observations. We considered two scenarios. In the first one, we assumed the DHOST model only as an alternative to DE, so that the internal dynamics of the clusters would be ruled by standard DM plus some possible ``external'' influence of the DHOST from cosmological scales. In the second one, instead, we assumed that the DHOST was playing both the role of DE and of DM in its entirety. In the former case, the DHOST model showed a mild Bayesian evidence over GR, although it alleviated the discrepancy present in GR between X-ray hydrostatic and lensing mass estimates. In the latter case, GR still seemed to be statistically mildly favored with respect to the DHOST model. 

Here we continue to explore the same DHOST model at galactic scales. In particular, we analyze the internal kinematics of a class of low surface brightness galaxies, called Ultra-diffuse galaxies (UDGs) \cite{vanDokkum:2014cea,2017ApJ...842..133L}. 
Quite recently one of them, NGC1052-DF2, has been under deep scrutiny because it seems to be characterized by an abnormally low amount of DM \cite{vanDokkum:2018vup,Pina:2021wkn}. It has become (in)famously known as the ``lacking DM galaxy'', although such strong statement might be deemed as too much overhasty, as suggested by the debate which has followed. Nevertheless, such deficiency in DM content must be explained somehow, and it has even been considered as a crucial deathblow to ETGs, because \textit{``$\ldots$a dark matter signature should always be detected, as it is an unavoidable consequence of the presence of ordinary matter''} \cite{vanDokkum:2018vup}. Thus, we think it is quite interesting to analyse it and verify if such claims are correct.

This lack of DM is indeed strange since DM plays a crucial role in the galaxy formation process, as galaxies are supposed to form due to the cooling and condensation of gas in potential wells of DM halos. In particular, UDGs have been detected primarily in highly dense environment \cite{2016IAUS..317...27M,vanDokkum:2014cea,2015ApJ...813L..15M}, but also in galaxy groups and even in voids \cite{2017MNRAS.468.4039R,2017ApJ...842..133L,2019MNRAS.486..823R,2019MNRAS.485.1036M}. Their presence in different environments demonstrates that their low DM content could be a consequence of the interaction with the surrounding environment \cite{2016MNRAS.459L..51A,2017MNRAS.466L...1D,Yang:2020iya,Maccio:2020xpw,Montes:2020zaz}, or due to specific internal processes \cite{10.1093/pasj/56.1.29,2020MNRAS.493.4700L,2013ApJ...775..116R}. Indeed, hydrodynamical simulations \cite{2018MNRAS.480L.106O,2018ApJ...863L..17N,Maccio:2020xpw} have shown that it is even possible to create a galaxy lacking DM in the standard cosmological scenario, and in \cite{Silk:2019jbt,2020ApJ...899...25S,Lee:2021wfv} it has been suggested that NGC1052-DF2 may have formed as a consequence of high velocity collisions of gas-rich galaxies. In a most recent work \cite{vanDokkum:2022zdd}, it was added that it might represent a sample of a possibly larger family, as a total of eleven low surface galaxies with the same properties have been found.

Further explanations regarding NGC1052-DF2 have been put forward. In \cite{2019MNRAS.486.1192T} it is claimed that NGC1052-DF2 might be located at a distance of $13$ Mpc, lower than the $\sim19$ Mpc from \cite{vanDokkum:2018vup}, allowing enough room for enough DM to be in full concordance with the standard picture and thus discarding any peculiarity. Other measurements \cite{Shen:2021zka} seem to confirm instead a larger distance, pushing it up to $22.1$ Mpc. 
In \cite{2019MNRAS.484..245L} it is claimed that the uncertainties on the mass estimates have been underestimated and they are actually much larger, so that only a weak inference can be performed. In \cite{2018MNRAS.481L..59H} it is shown how the choice of the tracer densities might have an influence. And in \cite{2018ApJ...859L...5M} the point is raised about how to analyze in the proper statistical way the observational data, as it might be that they are not really Gaussian distributed.

It is thus clear that the measurement of the dynamical mass of these objects from kinematical and dynamical data is crucial to discriminate between several formation processes and even to confute (or not) ETGs. Due to their low surface brightness, which implies a low signal-to-noise ratio, a measurement of the dynamical mass of UDGs through stellar velocity dispersion is not affordable. An alternative method, particularly well suited for UDGs, is based on the  dynamics of the globular clusters (GCs) within them. The poor-gas galaxy NGC1052-DF2 was observed with the Dragonfly Telescope Array \cite{2014PASP..126...55A}\footnote{The galaxy was already catalogued in \cite{2000A&AS..145..415K}.} and its dynamics was studied in detail in \cite{vanDokkum:2018vup,Wasserman:2018scp,Danieli:2019zyi}) exactly using GCs. From such observations it was inferred for the first time the possibility that it could host a very low amount of DM, pointing toward the possibility of a galactic dynamics mainly dominated by stars.

As anticipated above, the very low amount of DM shown by NGC1052-DF2 has been seen since the beginning as a critical problem for ETGs. Some attempts have been made to explain its dynamics within the Modified Newtonian Dynamics (MOND) scenario \cite{Milgrom:2014usa}. However, as specified in \cite{vanDokkum:2018vup}, within the MOND framework the expected velocity dispersion of NGC1052-DF2, derived using its population of GCs, appears to be a factor of two higher than the $90\%$ of the upper limit on the observed velocity dispersion. 

In \cite{Islam:2019szu,Islam:2019irh}, the dynamics of NGC1052-DF2 has been explored in many different ETGs scenarios. In addition to MOND, the authors considered MOND with External Field Effect (EFE) \cite{Hees:2015bna}, Weyl conformal gravity \cite{Mannheim:1988dj, Mannheim:2005bfa} and Moffat's Modified Gravity (MOG) \cite{Moffat:2005si}. In \cite{Islam:2019irh}, it has been shown that the MOG and the MOND with (EFE) \cite{Haghi:2019zpc,Famaey:2018yif} provide a good fit to the data, while the Weyl conformal gravity fits acceptably the data concluding that NGC1052-DF2 does not imply a dead-end for ETGs. 

For all the above reasons, we are interested here to study if the dynamics of NGC1052-DF2 is a trouble for the DHOST model we have chosen. As done in \cite{10.1093/mnras/stac180}, we will test the reliability of two different scenarios. In the first one, we will study the internal kinematics of NGC1052-DF2 with the DHOST model as an alternative to DE only. In the second scenario, we will test a unified scenario of DE and DM, where both have influence due to the partial breaking of the Vainshtein screening mechanism, with no need of adding any ad hoc DM component.

The paper is organized as follows: in Sec.~\ref{sec: Model} we introduce the DHOST model and the theoretical basis for the analysis of NGC1052-DF2 dynamics; in Sec.~\ref{sec: data} we present how we model the galaxy and each of its mass components; in Sec.~\ref{sec:stat} we explain how we performed our statistical analysis; in Sec.~\ref{sec: Results} we present the results and the implications of our analysis; in Sec.~\ref{sec: conclusions} we draw our conclusions.

\section{Theoretical Model}
\label{sec: Model}

The DHOST theory we consider here \cite{Crisostomi:2016czh, PhysRevD.97.021301, PhysRevD.97.101302} exhibits
a Vainshtein screening mechanism which is partially broken, leading to the gravitational ($\Phi$) and metric ($\Psi$) potentials:
\begin{align}
\label{eqn: model}
\frac{d\Phi}{dr} &= \frac{G_NM(r)}{r^2} + \Xi_1 G_N M''(r)\, , \\
\frac{d\Psi}{dr} &= \frac{G_NM(r)}{r^2} + \Xi_2 \frac{G_N M'(r)}{r} + \Xi_3 G_N M''(r)\, ,
\end{align}
where: $G_N$ is the measured gravitational constant that might be different from the bare one $G$ defined through the Planck Mass $M_{Pl} = (8\pi G)^{-1}$; $M(r)$ is the spherical mass enclosed in the radius $r$; $M'(r)$ and $M''(r)$ are respectively the first and second order derivative of the mass with respect to the radius $r$; and $\Xi_{1,2,3}$ (using the notation of \cite{Cardone:2020rmy}) are the three coupling parameters that characterize the model (they all go to zero in the GR limit). Since we are not interested in a cosmological analysis of the DHOST model, through this work we assume that it reproduces fairly well a $\Lambda$CDM background cosmology with $H_0 =67.74$ km s$^{-1}$Mpc$^{-1}$, $\Omega_m =0.3089$ and $\Omega_\Lambda = 0.6911$ \cite{Planck:2018vyg}.

The parameters $\Xi_{1,2,3}$ can be written in terms of more fundamental Effective Field Theory (EFT) constants \citep{Dima:2017pwp}. Taking into account the constraints which can be derived from multi-messenger observation of GW170817 \citep{LIGOScientific:2017ync}, such relations can be expressed as \citep{Creminelli:2017sry,Sakstein:2017xjx,Ezquiaga:2017ekz,Baker:2017hug,Baker:2020apq}:
\begin{eqnarray}
\label{eqn: xi param}
\Xi_1 &=& -\frac{1}{2}\frac{(\alpha_H+\beta_1)^2}{\alpha_H+2\beta_1}\, ,\\
\Xi_2 &=& \alpha_H\, ,\\
\Xi_3 &=& -\frac{\beta_1}{2}\frac{(\alpha_H+\beta_1)}{\alpha_H+2\beta_1}\, , \\
\gamma_0 &=& - \alpha_H - 3 \beta_1\, ,
\end{eqnarray}
where $\gamma_0$ is the fractional difference between $G$ and $G_N$,
\begin{equation}\label{eq:gamma0}
\gamma_0 = \left(8 \pi M_{PL}^{2} G_N \right)^{-1}-1\, .
\end{equation}
The EFT parameters are: $\alpha_H$, which is connected to the kinetic mixing between matter and the scalar field introduced by the DHOST theory \citep{Dima:2017pwp}; and $\beta_1$, which parameterizes the presence of higher-order operators in the lagrangian \citep{Dima:2017pwp}. Note also that we assume $G_N$ fixed at its measured value, so that $\gamma_0$ is fully determined by $\alpha_H$ and $\beta_1$, while $G$ should be derived from them.

\subsection{Galactic Dynamics theory}

The internal kinematics of NGC1052-DF2 can be described using the Jeans equation (with the assumptions of spherical symmetry and collisionless tracers population):
\be
\label{eqn: Jeans}
\frac{d(l(r)\sigma_r(r))^2}{dr} + \frac{\beta(r)}{r}(l(r)\sigma_r(r))^2 = l(r)\frac{d\Phi(r)}{dr}\, ,
\ee
where $l(r)$ is the luminosity density of the galaxy and $\beta(r)$ is the anisotropy parameter \cite{2008gady.book.....B},
\be
\label{eqn: anisoptropy}
\beta(r) = 1 - \frac{\sigma^2_t(r)}{\sigma^2_r(r)}\, ,
\ee
where $\sigma_t$ is the tangential velocity dispersion (defined as a combination of the two angular components of the velocity dispersion tensor $\sigma^2_t = (\sigma^2_\theta + \sigma^2_\phi)/2$), and $\sigma_r$ is the radial component of the velocity dispersion tensor. If $\beta = 0$ the system is fully isotropic; if $\beta = 1$ the system is defined as purely radial; if $\beta \to -\infty$ the system is purely tangential.

The information about the gravity theory is fully encoded in the gravitational potential $\Phi(r)$ on the right hand side of Eq.~\eqref{eqn: Jeans}. Consequently, it is possible to study the internal kinematics of a galaxy using in our modified gravity scenario including Eq.~\eqref{eqn: model} in Eq.~\eqref{eqn: Jeans}.

In GR, integrating once equation Eq.~\eqref{eqn: Jeans} allows us to solve the Jeans equation finding \cite{Mamon:2004xk}
\be
\label{eqn: solution}
l(r)\sigma^2_r(r) = \frac{1}{f(r)}\int^\infty_rds\hspace{.1em}f(s)l(s)\frac{M(s)}{s^2}\, ,
\ee
with
\be
\label{eqn: effe}
\frac{d\log f(r)}{d\log r} = 2\beta(r)\, ,
\ee
where $f(r)$ depends on the specific parametrization used for the anisotropy parameter $\beta(r)$. Projecting Eq.~\eqref{eqn: solution} along the line of sight, we define the line-of-sight (\textit{los}) velocity dispersion as
\begin{align}
\label{eqn: vlos}
\sigma^2_{los}(R) = &\frac{2}{I(R)}\int^\infty_R dr\,r\frac{l(r)\sigma^2_r(r)}{\sqrt{r^2 - R^2}} - \nonumber\\
&R^2\int^\infty_R dr\,\beta(r)\frac{l(r)\sigma^2_r(r)}{r\sqrt{r^2 - R^2}}\, ,
\end{align}
where $R$ is the projected radius, and $I(R)$ is the stellar surface brightness. 

Finally, inserting Eq.~\eqref{eqn: solution} into Eq.~\eqref{eqn: vlos}, one can obtain the velocity dispersion along the line as
\be
\label{eqn: finvlos}
\sigma^2_{los}(R) = \frac{2G_N}{I(R)}\int^\infty_Rdr\hspace{.1em}K\biggl(\frac{r}{R}\biggr)l(r)M(r)\frac{dr}{r}
\ee
where $K(r/R)$ is the kernel function whose expression depends on the specific parametrization of the anisotropy parameter $\beta(r)$ (see \cite{Mamon:2004xk} for more details). 

In order to generalize Eq.~(\ref{eqn: finvlos}) to our DHOST scenario, we notice that Eq.~(\ref{eqn: model}) can be written as
\begin{align}\label{eqn: model_eff}
\frac{d\Phi}{dr} &= \frac{G_NM_{eff}(r)}{r^2}\, ,
\end{align}
where we define the effective mass as $M_{eff}(r) = M(r) + \Xi_1 r^2 M''(r)$. In such a way, Eq.~(\ref{eqn: finvlos}) simply becomes
\be
\label{eqn: finvlos_DHOST}
\sigma^2_{los}(R) = \frac{2G_N}{I(R)}\int^\infty_Rdr\hspace{.1em}K\biggl(\frac{r}{R}\biggr)l(r)M_{eff}(r)\frac{dr}{r}\, .
\ee
Note also that by using Eq.~(\ref{eqn: finvlos_DHOST}) we can put direct constraints on $\Xi_1$ only, which in turn will result on indirect correlated constraints on $\{\alpha_{H}, \beta_1\}$.

\section{Data and Galaxy model}
\label{sec: data}

The UDG NGC1052-DF2 was initially identified using the Dragonfly Telescope Array in \cite{Abraham:2014lfa}.
Structural photometric parameters, like the size, surface brightness, magnitude and color, have been measured in combination with the \textit{Hubble} and reported in \cite{vanDokkum:2018vup}. Here we use such data\footnote{Data are taken from \url{http://www.astro.yale.edu/dokkum/outgoing/ascii_table.txt}, with one revised velocity from \cite{2018RNAAS...2...54V}.}. 

The spectroscopy of the compact objects associated with the galaxy was realized with the W. M. Keck Observatory in two different runs. The first one was carried out using the Deep Imaging Multi-Object Spectrograph (DEIMOS) \cite{2003SPIE.4841.1657F} on Keck II, and the last one with the Low-Resolution Imaging Spectrometer (LRIS) \cite{1995PASP..107..375O} Keck I. Ten compact objects similar to globular clusters have been observed, for which both radial velocities and velocity dispersion could be measured \cite{vanDokkum:2018vup}. 

The dynamics of NGC1052-DF2 has been studied in \cite{Wasserman:2018scp,Danieli:2019zyi,Shen:2021zka,2019A&A...625A..77F}. In the following sections we describe our galaxy modelling and the differences between our analysis and the literature.

\subsection{Stellar component}

The stellar component of the galaxy is modeled as a single S\'{e}rsic profile
\be
\label{eqn: lum}
I(R) = I_{0} \exp \left[ -\left( \frac{R}{a_{s}}\right)^{1/n}\right]\,,
\ee
with: $I_0$, the central surface brightness; $a_s$, the S\'{e}rsic scale parameter; $n$, the S\'{e}rsic index. The S\'{e}rsic scale parameter $a_s$ is generally expressed in terms of the effective (half-light) radius $R_{eff}$ by the relation
\be
a_s = \frac{R_{eff}}{(b_n)^n}
\ee
with $b_n = 2n - 0.33$ \cite{1993MNRAS.265.1013C}. The photometric analysis presented in \cite{vanDokkum:2018vup} fixes the S\'{e}rsic index $n = 0.6$, the effective radius $R_{eff} = 22.6$ arcsec and $I_0 = 24.4$ mag arcsec$^{-2}$ in the $V_{606}$ band. At the fiducial distance $D = 20$ Mpc of \cite{vanDokkum:2018vup}, this corresponds to an absolute magnitude $M_V= -15.4$ and luminosity (in solar units) $L_{tot} = 1.12\cdot 10^8 L_\odot$.

The luminosity density could be derived by Abel inversion of Eq.~\eqref{eqn: lum}, but no analytical expression is known for a S\'{e}rsic with free $n$. In \cite{1997A&A...321..111P} it is shown that it can be well approximated by a function 
\be
\label{eqn: lumin}
\ell(r) =\ell_1\widetilde{\ell}(r/a_s)\, ,
\ee
with:
\begin{equation}
\widetilde{\ell}(x) \simeq x^{-p_n} \, \exp(-x^{1/n}) \; ,
\end{equation}
\begin{equation}
\ell_{1} = \frac{L_\mathrm{tot}}{4 \pi \, n \, \Gamma[(3-p_n)n] a_{s}^{3}} \; .
\end{equation}
The function $p_n$ is defined in \cite{Neto:1999gx} as:
\begin{equation}
p_n \simeq 1.0 - 0.6097/n + 0.05463/n^2 \; ,
\end{equation}
and the total galaxy luminosity is
\begin{equation}
\label{eq:ltot}
L_{tot} = 10^{-0.4(m_{V_{606}} - \mu(D) - M_{\odot,V_{606}})}
\end{equation}
where $m_{V_{606}}$ is the apparent magnitude of the galaxy; $\mu(D) = 5\log_{10}D + 25$ is the distance modulus, with the distance to the galaxy, $D$, expressed in Mpc; and $M_{\odot,V_{606}}$ is the total magnitude of the Sun in the $V_{606}$ band \cite{2018ApJS..236...47W}.

To get the mass density we have to multiply Eq.~\eqref{eqn: lumin} by the light-to-mass ratio $\Upsilon$,
\be
\label{eqn: stardens}
\rho_*(r) = \Upsilon\ell\biggl(\frac{r}{R_e}\biggr)\, .
\ee
Finally, the total mass enclosed in the radius $r$ can be derived by integrating Eq.~(\ref{eqn: stardens}), and can be expressed analytically as
\begin{align} \label{eqn: stellar mass}
&M_*(<r) = 2\pi n \Upsilon I_0 \left(\frac{R_{eff}}{b^n_n}\right)^2\frac{\Gamma(2n)}{\Gamma[(3 - p_n)n]} \times\\
&\left\{\Gamma[(3 - p_n)n] - \gamma\left[(3 - p_n)n,b_n\left(\frac{r}{R_{eff}}\right)^{1/n}\right]\right\}\, ,
\end{align}
where $\Gamma$ and $\gamma$ are the Euler and the incomplete gamma functions respectively.

As it is well known, Eq.~\eqref{eqn: Jeans} is characterized by a degeneracy between the radial dispersion profile and the velocity anisotropy, which is practically not measurable. This requires assumptions on the anisotropy parameter $\beta(r)$ in order to correctly recover the mass profile of the galaxy. We consider two different possibilities. The first case is a constant anisotropy profile, $\beta(r)\equiv \beta_c$. Then, we also analyze a model with a radial anisotropy profile, first proposed in \cite{Mamon:2012yb} and then found specifically appropriate for UDGs in \cite{Zhang:2015pca}, given by
\be
\label{eqn: anis}
\beta(r) = \beta_0 + (\beta_\infty - \beta_0)\frac{r}{r + r_a}\, ,
\ee
where $\beta_0$ is the inner anisotropy (at $r = 0$); $\beta_\infty$ is the outer anisotropy (at $r = \infty$); and $r_a$ is the scale radius of profile. This radial anisotropy functional form is a monotonic increasing function that reduces to a constant anisotropy $\beta\equiv\beta_c = \beta_0$ if and only if the inner and the outer anisotropy contributes coincide. Each form of $\beta(r)$ implies in Eq.~\eqref{eqn: finvlos} a specific Kernel function, as tabulated in \cite{Mamon:2004xk}.

Thus, in our analysis the free parameters for the stellar component will be $\{\Upsilon, D, \beta_c\}$ or $\{\Upsilon, D, \beta_0, \beta_{\infty}, r_a\}$ depending on the considered velocity dispersion model. It is worth to stress here that a change in the distance $D$ introduces a change in the conversion factor between arcsec and kpc, thus affecting both the values of the S\'{e}rsic parameters $R_{eff}$ and $I_{0}$, when expressed in kpc and solar units, respectively, and of the data of the dispersion curve, which in \cite{vanDokkum:2018vup} is provided in kpc (at their fiducial distance).

%For the computation of the Kernel, we have to distinguish, respectively, the cases of $u_a \neq 1$ and $u_a = 1$ (\cite{Mamon:2004xk} for the expressions of the kernels).

\subsection{Dark matter component}

When we assume a DM component, we consider a generalized Navarro-Frenk-White density profile (gNFW) \cite{Zhao:1995cp,Kravtsov:1997dp,Moore:1999gc} 
\be
\label{eqn: gNFW}
\rho_{gNFW}(r) = \rho_s\biggl(\frac{r}{r_s}\biggr)^{-\gamma}\biggl(1 + \frac{r}{r_s}\biggr)^{\gamma - 3}
\ee
where $\rho_s$ and $r_s$ are the characteristics NFW density and radius parameters and $\gamma$ represents the inner log-slope. For $\gamma = 1$ we recover the typical NFW density profile of DM halos \cite{Navarro:1995iw,Navarro:1996gj}, but Eq.~(\ref{eqn: gNFW}) is preferred because it has more freedom and (if needed) can recover a larger variety of inner profiles than the standard NFW profile. 

From Eq.~\eqref{eqn: gNFW} we can derive the mass profile, as
\begin{align}
M_{DM}(<r) &= \frac{4 \pi \rho_{s} r^{3}_{s}}{3-\gamma} \left(\frac{r}{r_{s}} \right)^{3-\gamma}\, \\ 
&_2F_1[3 - \gamma,3- \gamma,4 - \gamma,- \frac{r}{r_s}]\, , \nonumber
\end{align}
with $_2F_1$ being the hypergeometric function.

When working with DM, it is common to introduce the concentration parameter, $c_{\Delta}$, defined as
\be
c_{\Delta} = \frac{r_{\Delta}}{r_s}\, ,
\ee
where the $\Delta$ means that all quantities are calculated at the radius $r_{\Delta}$, where the density of the system is $\Delta$ times the critical density of the Universe, $\rho_c(z)$, at the same redshift of the object. In our case, we consider $\Delta=200$, the so-called virial value, and the critical density of the Universe reads,
\begin{equation}
\rho_{c}(z) = \frac{3 H^{2}(z)}{8\pi G_{N}}\, ,
\end{equation}
where $H(z)$ is from the fiducial $\Lambda$CDM cosmology we have defined in Sec.~\ref{sec: Model}, and $z= 0.004963$ is the redshift of NGC1052 derived from the NED database\footnote{\url{https://ned.ipac.caltech.edu/}}.

Thus, we can introduce the virial mass, $M_{200}$ 
\be
M_{200} = \frac{4\pi}{3}200\rho_cr^3_{200}\, ,
\ee
from which we can express the scale density $\rho_s$ as
\be
\label{eqn: rhos}
\rho_s = \frac{200}{3}\rho_{c}(z) \frac{(3-\gamma)(c_{200})^\gamma}{_2F_1[3 - \gamma,3- \gamma,4 - \gamma,- c_{200}]}\, .
\ee
For convenience, the free parameters for the gNFW component in our statistical analysis will be $\{c_{200},M_{200},\gamma\}$.

\section{Statistical analysis}
\label{sec:stat}

Assuming a Gaussian likelihood, the $\chi^2$ for the data we have is defined as 
\begin{equation}
\chi^2(\boldsymbol{\theta}) = \sum_{i}^{\mathcal{N}_{data}}  \frac{(v_i - v_{sys})^2}{\sigma_{i}^2} + \ln \left(2\pi \sigma_{i}^2\right)
\end{equation}
where: $\mathcal{N}_{data}=10$ is the number of observed GCs within NGC1052-DF2; $v_{sys}$ is the systemic velocity of NGC1052-DF2 (and thus of the ten GCs); $v_{i}$ are the observed velocities of the ten GCs (at given distances from the center of the galaxy, $R_{i}$); $\sigma_{i}^2 = \sigma^2_{los,i}(\boldsymbol{\theta}) + \sigma^2_{v_i}$ is the total error budget on the velocities $v_{i}$, with $\sigma_{v_{i}}$ the measurement uncertainty and $\sigma^2_{los,i}(\boldsymbol{\theta})$ the velocity dispersion, which explicitly depends on the model parameters (see Eqs.~\eqref{eqn: finvlos} and \eqref{eqn: finvlos_DHOST}); and $\boldsymbol{\theta}$ is the vector of the model parameters. When we work with GR and with a constant anisotropy profile, $\boldsymbol{\theta} = \{v_{sys},D,\Upsilon,\beta_c,c_{200_c},M_{200},\gamma\}$; instead with a radial anisotropy, given by Eq.~\eqref{eqn: anis},  $\boldsymbol{\theta} = \{v_{sys},D,\Upsilon,\beta_0,\beta_\infty,r_a,c_{200_c},M_{200},\gamma\}$. When the DHOST model is considered as DE only we have $\boldsymbol{\theta} = \{v_{sys},D,\Upsilon,\beta_c,c_{200_c},M_{200},\gamma,\Xi_1\}$ taking a constant anisotropy parameter and $\boldsymbol{\theta} = \{v_{sys},D,\Upsilon,\beta_0,\beta_{\infty},r_a,c_{200_c},M_{200},\gamma,\Xi_1\}$ with a radial one. When the DHOST model is both DE and DM, we have $\boldsymbol{\theta} = \{v_{sys},D,\Upsilon,\beta_c,\Xi_1\}$ and $\boldsymbol{\theta} = \{v_{sys},D,\Upsilon,\beta_0,\beta_{\infty},r_a,\Xi_1\}$.

Additionally, we also apply one control and a series of priors. The control is of physical nature: $\sigma_{los}>0$ at any distance $R$ from the center which is sampled by the data. This control is designed to check if the MCMC explores a region of the parameters which, even if satisfying the priors which we define in the following, may still return an unphysical value for the velocity dispersion. 

Moreover, we apply: on the systemic velocity, a Gaussian prior $v_{sys} = 1801.6\pm5$ km s$^{-1}$ \cite{Wasserman:2018scp}; on the mass-to-light ratio, another Gaussian prior $\Upsilon = 1.7 \pm 0.5$, derived from stellar population studies \cite{vanDokkum:2018vup,2018ApJ...856L..30V}; while on the dispersion velocity parameters, we have a lognormal prior $\log \left(1-\beta_{i}\right) = 0 \pm 0.5$ km s$^{-2}$ (where $\beta_{i} = \{\beta_c,\beta_0,\beta_{\infty}\}$) on the range $\beta_{i}\in[-10,1]$ \cite{Wasserman:2018scp}.

\subsection{Prior on the distance}

The distance of the galaxy NGC1052-DF2 is a sensitive element in the analysis of this galaxy, as it has been envisaged \cite{2019MNRAS.486.1192T} that a different (closer) location might resolve at all the strangeness of a galaxy lacking dark matter.

In \cite{vanDokkum:2018vup} the distance is estimated by the Surface Brightness Fluctuations (SBF) method \cite{1988AJ.....96..807T,Blakeslee:2010dn} and is $D= 19.0 \pm 1.7$ Mpc. In \cite{Wasserman:2018scp} a compatible result always derived from SBF is $D = 19.0 \pm 1$ Mpc. In \cite{Shen:2021zka}, the distance is estimated using the method of the tip of the red-giant branch (TRGB) \cite{1993ApJ...417..553L}. Measurements with the Advance Camera for Surveys (ACS) of \textit{Hubble} give a result of $D = 22.1 \pm 1.2 $ Mpc. This measurement completely rules out the one from \cite{2019MNRAS.486.1192T}, of $D = 13$ Mpc, and even states that the ten GCs objects are more luminous than previously measured. This is, eventually, the prior we apply on the distance $D$.

\subsection{Priors on \texorpdfstring{$c_{200}$}{c200} and \texorpdfstring{$M_{200}$}{M200}}

Given the limited extension (in distance from the center of NGC1052-DF2) of the GCs data, to leave the DM parameters $c_{200}$ and $M_{200}$ totally free, with only some uninformative flat prior on them, results in having such parameters totally unconstrained. For such a reason, the fit is performed applying a lognormal prior on $c_{200}$, taking advantage of the many $c-M$ relations which can be found in literature.

One of the most used is from \cite{Dutton:2014xda}, but it does not really cover the mass range of the UDGs in which we are more interested. In \cite{Wasserman:2018scp} the authors use the $c-M$ relation from \cite{Diemer:2018vmz}, which requires the calculation of too many cosmologically-related quantities, and we want to avoid this as it would imply to assume too many things both at the background and at the perturbative level, when we do not really have a fully detailed literature about DHOST in the context of matter power spectrum and growth of perturbations. We must point out also that all the $c-M$ relations available in literature are based on GR simulations. Thus, any usage of them in an ETG context is somehow extrapolated.

In this work we have chosen the relation provided by \cite{Correa:2015dva}, which separately considers different redshifts ranges and is updated to a \textit{Planck} 2015 cosmology (see their Appendix B1).
The standard deviation we apply is $\sigma_{\log c_{200}} = 0.16$ dex.  

We additionally consider two different scenarios for the prior on the halo mass $\log M_{200}$: one lognormal prior based on the stellar-to-halo mass relation (SHMR) from \cite{2017MNRAS.470..651R}, which constrains $M_{200}$ from the total stellar mass $M_{\ast} = \Upsilon L_{tot}$, with dispersion $0.3$ dex; and one with no SHMR, in which we have an uniformative flat prior on $\log M_{200}\in[2,15]$. For the gNFW slope parameter $\gamma$, we consider a uniform flat prior on the range $[0,2]$.

Finally, it is important to stress that when we work in the scenario in which the DHOST might play the role of an effective DM component, we do not apply any prior at all on such possible effective-DM behaviour, and we leave the parameter $\Xi_1$ totally free.

\subsection{Bayesian analysis}

The total $\chi^2$, taking into account all the priors, is minimized using our own Monte Carlo Markov Chain (MCMC) code, whose convergence is checked using the method developed in \cite{Dunkley:2004sv}.

After having performed the fit, the next most important step is to compare the GR and the DHOST scenario in a reliable way. For this, we compute a series of quantities which are recognized nowadays as the most accurate for this goal \cite{Abdalla:2022yfr}. Taking advantage of the outputs of our MCMCs, which directly provide the posteriors for each parameters, we use our own implementation of the nested sampling algorithm described in \cite{2006ApJ...638L..51M} to calculate them with the appropriate modifications.

First of all, we calculate the evidence $\mathcal{E}_{i} \equiv \mathcal{E}(\mathcal{M}_{i})$, defined as the probability of the data $\mathbf{d}$ given the model $\mathcal{M}_{i}$ with a set of parameters $\boldsymbol{\theta}$,
\begin{equation}\label{eq:evidence}
\mathcal{E}_{i} = \int d\boldsymbol{\theta}\, \mathcal{L}_{i}(\boldsymbol{\theta}) \pi_{i}(\boldsymbol{\theta}) \, ,
\end{equation}
where $\mathcal{L}_{i} \propto \exp^{-\chi_{i}^2}$ and $\pi_{i}$ are respectively the likelihood and the prior function of the model $\mathcal{M}_{i}$, with the posterior being $\mathcal{P}_{i} = \mathcal{L}_{i} \pi_{i}/\mathcal{E}_{i}$.

\begin{figure*}
\centering
\includegraphics[width=8.cm]{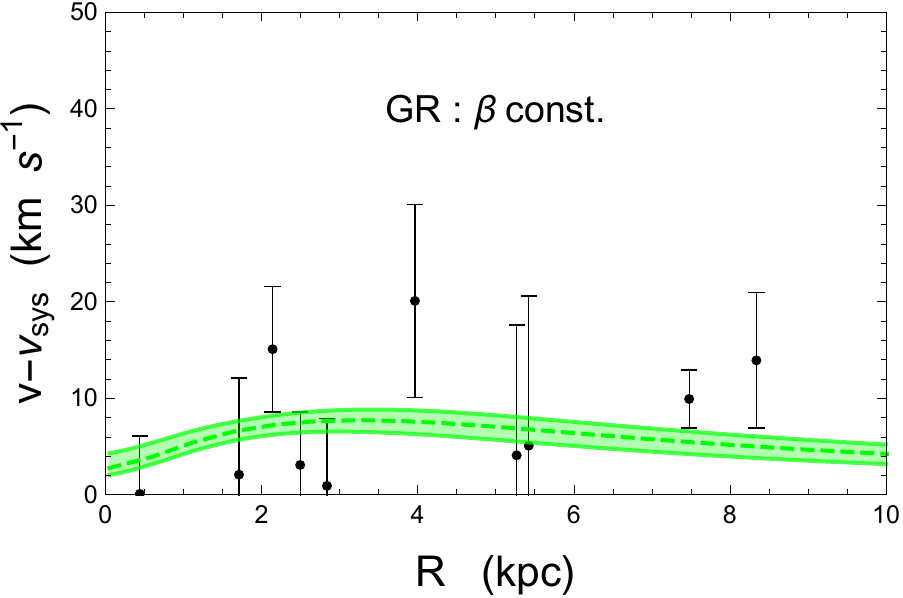}~~~
\includegraphics[width=8.cm]{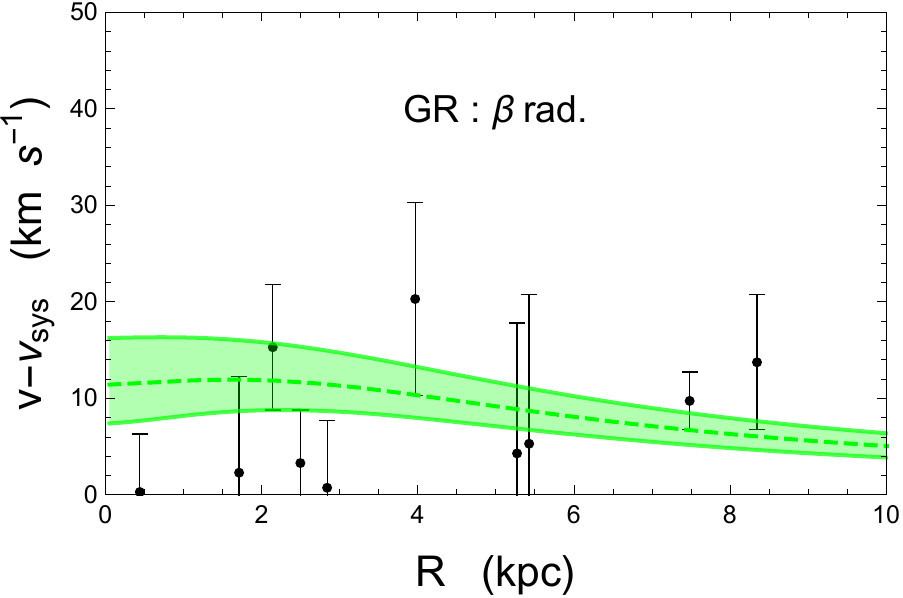}\\
\caption{Velocity offset profiles of NGC1052-DF2 with GR and no DM component, but only the stellar contribution. Black dots and bars are observational data, $v_{i}-v_{sys}$, with uncertainties, $\sigma_{v_{i}}$. Colored dashed lines and shaded regions are respectively the median and the $1\sigma$ confidence region of the $\sigma_{los}$ profile derived from Eq.~\eqref{eqn: finvlos}. Left panel: constant velocity  anisotropy profile. Right panel: radial velocity profile.}\label{fig:plot_GR_STAR}
\end{figure*}

From it, we determine the Bayes Factor $\mathcal{B}^{i}_{j} \equiv \mathcal{E}_{i} / \mathcal{E}_{j}$, where our reference model $\mathcal{M}_j$ is the GR scenario with only stellar component. We interpret the Bayes Factor using the Jeffrey's scale \cite{Jeffreys1939-JEFTOP-5}: if $\ln \mathcal{B}^{i}_{j} < 1$ the evidence in favour of the model $\mathcal{M}_i$ is not significant respect to the one of $\mathcal{M}_j$; if $1 < \ln \mathcal{B}^{i}_{j} < 2.5$ the evidence of $\mathcal{M}_i$ is mild; if $2.5 < \ln \mathcal{B}^{i}_{j} < 5$ the evidence is strong; when $\ln \mathcal{B}^{i}_{j} > 5$ the evidence is decisive.

However, since the Bayes factor can be prior-dependent \cite{Nesseris:2012cq}, in order to perform a proper comparison between models, prior-independent quantities should be used. In \cite{Handley:2019wlz,Handley:2019pqx,Joachimi:2021ffv} a new statistical quantity called suspiciousness $\mathcal{S}^{i}_{j}$ is introduced and defined as
\be
\label{eqn: suspiciousness}
\log\mathcal{S}^{i}_{j} = \log\mathcal{B}^{i}_{j} +
\mathcal{D}_{KL,i} - \mathcal{D}_{KL,j}\,
\ee
to quantify the mismatch between the model $\mathcal{M}_i$ and $\mathcal{M}_j$ in a prior-independent way. The suspiciousness may be considered as the value of the Bayes ratio which corresponds to the narrowest possible priors that do not significantly alter the shape of the posteriors.

In Eq.~\eqref{eqn: suspiciousness}, $\mathcal{D}_{KL,i}$ is the Kullback-Leibler (KL) divergence \cite{10.1214/aoms/1177729694} which quantifies the information gain between prior and posterior,
\be
\mathcal{D}_{KL,i} = \int d\boldsymbol{\theta}\, \frac{\mathcal{L}_{i}(\boldsymbol{\theta})}{\mathcal{E}_{i}}
\log\frac{\mathcal{L}_{i}(\boldsymbol{\theta})}{\mathcal{E}_{i}}\, .
\ee
From the above definition, we can see that the KL divergence $\mathcal{D}_{KL,i}$ is prior-dependent too. But in \cite{Handley:2019pqx} it is shown how the Bayes ratio $\mathcal{B}^{i}_{j}$ and the difference between the KL divergences for $\mathcal{M}_i$ and $\mathcal{M}_j$ transform similarly by varying the prior volume, and therefore how in the suspiciousness $\mathcal{S}^{i}_{j}$ the prior dependency is removed.

The interpretation of the suspiciousness for model selection, in analogy to the Jeffrey scale, is described in Fig. 4 of \cite{Handley:2019pqx}: a consistency between two models $\mathcal{M}_i$ and $\mathcal{M}_j$ is achieved for positive values of $\log\mathcal{S}^{i}_{j}$, whilst a negative value of $\log\mathcal{S}^{i}_{j}$ shows a tension between the models. More specifically, \textit{``an overly negative value of $\log S^{i}_{j}$ indicates discordance, and an overly positive value suspicious concordance''} \cite{Handley:2019pqx,Handley:2019wlz}.

\section{Results}
\label{sec: Results}

All the scenarios which we have considered, depending on the mass components involved and on the priors applied, are shown in Table~\ref{tab:results}.

The first step has been to analyze the GR scenario, which is our reference model against which we can compare and assess the statistical validity of the DHOST model described by Eq.~\eqref{eqn: model}. More specifically, our reference scenario is the GR case with only stellar contribution and constant anisotropy profile. Note that in \cite{Wasserman:2018scp} there is always a DM component, so the scenario with only stars is described here for the first time. In the left panel of Fig.~\ref{fig:plot_GR_STAR} we can easily see, even just by visual inspection, how we can perfectly fit the data with only the stars, with a spatially averaged total dispersion $\sigma \sim 10.53$ km s$^{-1}$. From Table~\ref{tab:results} we see that no deviation from the applied priors is found, and the constant anisotropy profile shows a preference for highly tangential modes.

\begin{figure*}
\centering
\includegraphics[width=8.cm]{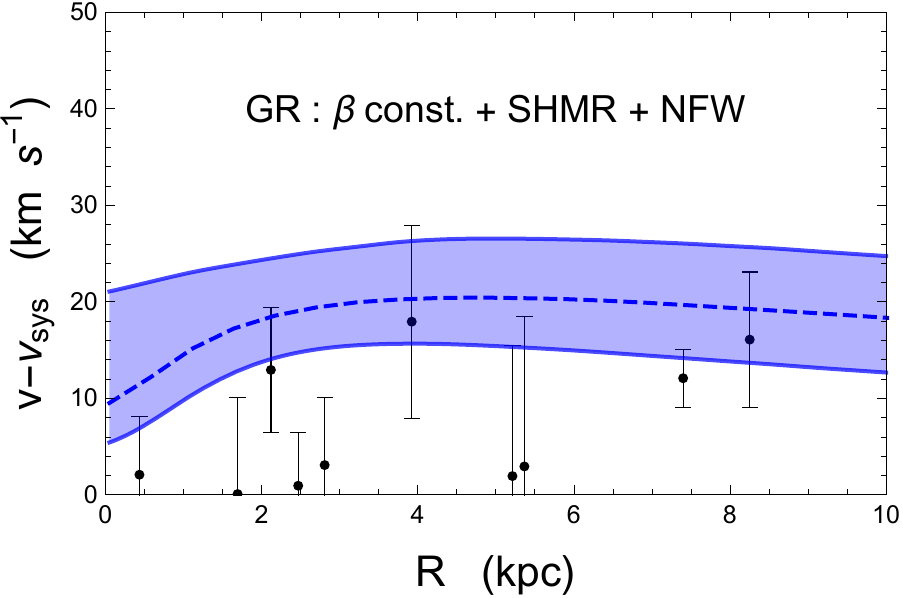}~~~
\includegraphics[width=8.cm]{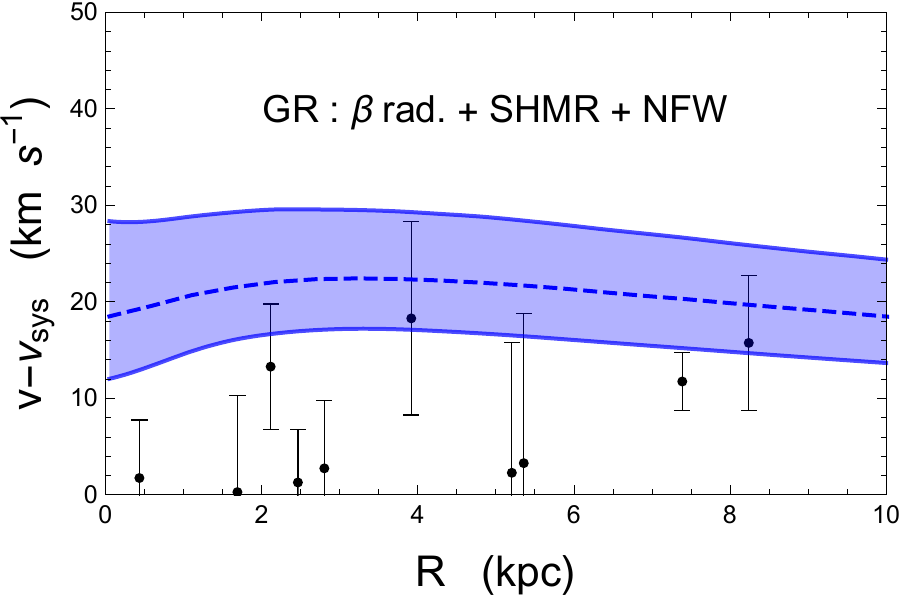}\\
~~~\\
\includegraphics[width=8.cm]{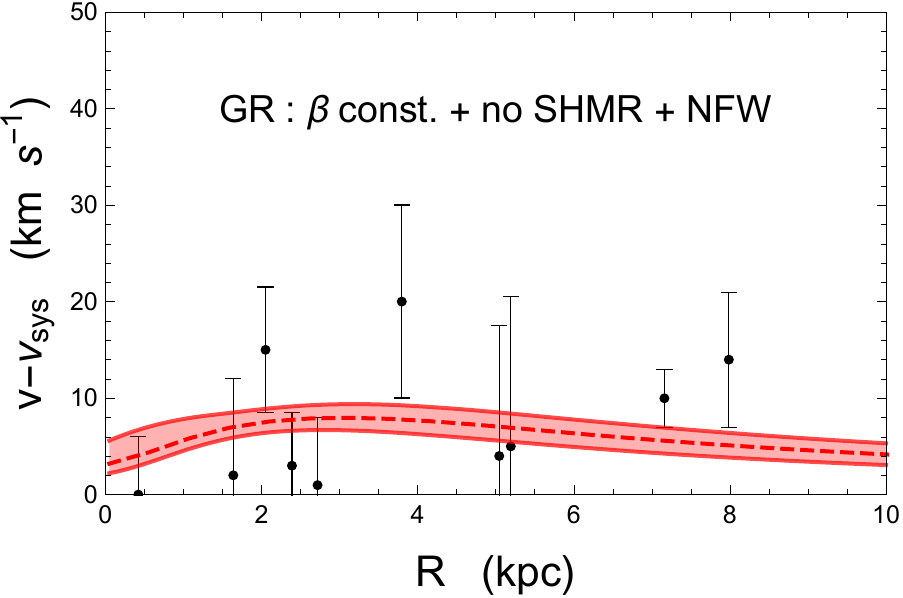}~~~
\includegraphics[width=8.cm]{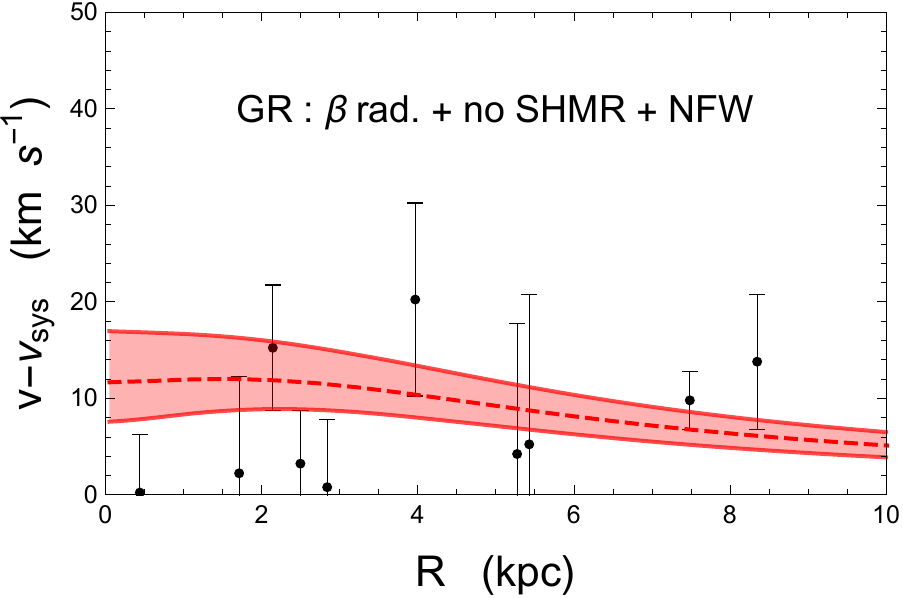}\\
\caption{Velocity offset profiles of NGC1052-DF2 with GR and a gNFW DM component. Black dots and bars are observational data, $v_{i}-v_{sys}$, with uncertainties, $\sigma_{v_{i}}$. Colored dashed line and shaded regions are respectively the median and the $1\sigma$ confidence region of the $\sigma_{los}$ profile derived from Eq.~\eqref{eqn: finvlos}. Left panels: constant velocity  anisotropy profile. Right panels: radial velocity profile. Top panels: with SHMR prior. Bottom panels: without SHMR prior.}\label{fig:plot_GR_NFW} 
\end{figure*}

In the right panel of Fig.~\ref{fig:plot_GR_STAR} we have the same scenario, but we allow for a radial variation of the anisotropy function. From Table~\ref{tab:results} we can see how there is no statistically significant difference with respect to the previous case. We detect a rise in the velocity offset at small scales, while at large scales we have the same descending profile, although the spatially average dispersion velocity is still consistent with the data, being $\sigma \sim 11.10$ km s$^{-1}$.  The $\beta$ parameters are quite well constrained, still pointing to a highly tangential anisotropy profile, but we also note how the radius parameter $r_a$ is basically unconstrained. Moreover, we see that from the Bayesian point of view, this scenario is disfavoured with respect to our chosen reference one: not in a significant way for the Jeffrey's scale, but already in a ``statistical tension'' regime, although very mild, when looking at the suspiciousness.

When we include a gNFW DM component in the GR scenario, things change drastically and are strongly related to the presence of the SHMR prior. All cases are shown in Fig.~\ref{fig:plot_GR_NFW}. The first main differences among the SHMR and the no-SHMR case are a slight shift toward a smaller distance, a smaller mass-to-light ratio and a lower systemic velocity in the former case, although they are still statistically consistent with each other. Moreover, the anisotropy parameters are less negative, but still fully tangential. In the SHMR case, the spatially averaged dispersion is $\sigma \sim 17$ km s$^{-1}$, thus in tension with most of the data; while in the no-SHMR case we have $\sigma \sim 11$ km s$^{-1}$, much more in agreement with observations.

When looking in more details to the DM parameters, we see the most net differences. The SHMR case has perfect gaussian constraints on both $c_{200}$ and $M_{200}$, while for $\gamma$ we can only set an upper limit, being it basically consistent with zero. We must point out that the median value we get for $c_{200}\sim 8$, does not exactly correspond to the median value we would expect from the $c-M$ relation from \cite{Correa:2015dva} using our $M_{200}$ MCMC outputs, which would be $c_{200}\sim11$. Although, given the uncertainties, they are still statistically consistent with each other. Nevertheless, the SHMR case is both strongly disfavoured and in strong tension with the reference scenario. 

Things are different when we relax the SHMR assumption, and we do not impose such a prior. All parameters are perfectly consistent with the reference case, but now the DM has quite different properties: a very high concentration, $c_{200}\sim 24$; a very low mass content, $\log M_{200}<7$ $M_{\odot}$; and $\gamma$ totally unconstrained, with a uniform distribution all over the allowed range. Note that the limit on $M_{200}$ is only an upper one.

\begin{figure*}
\centering
\includegraphics[width=8.5cm]{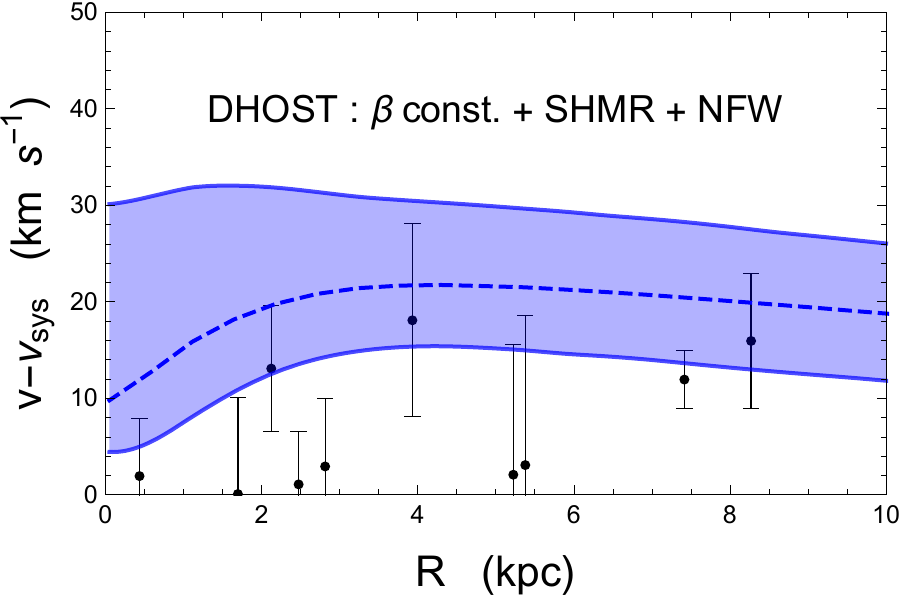}~~~
\includegraphics[width=8.5cm]{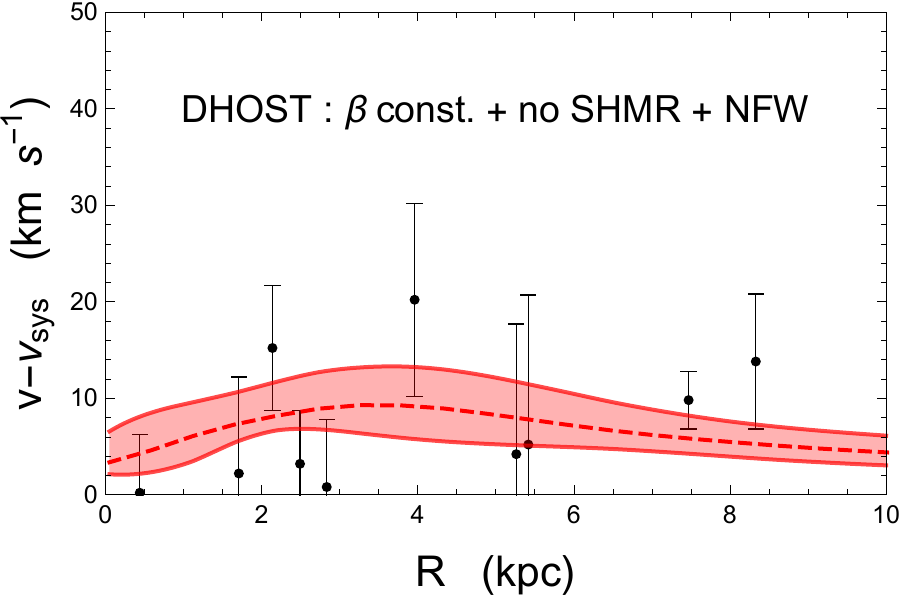}\\
~~~\\
\includegraphics[width=8.5cm]{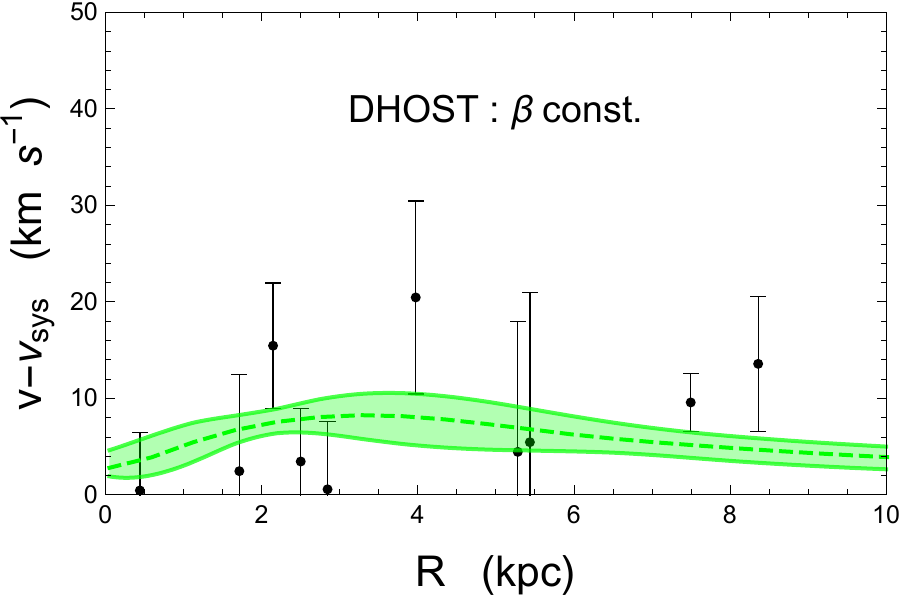}~~~
\includegraphics[width=8.5cm]{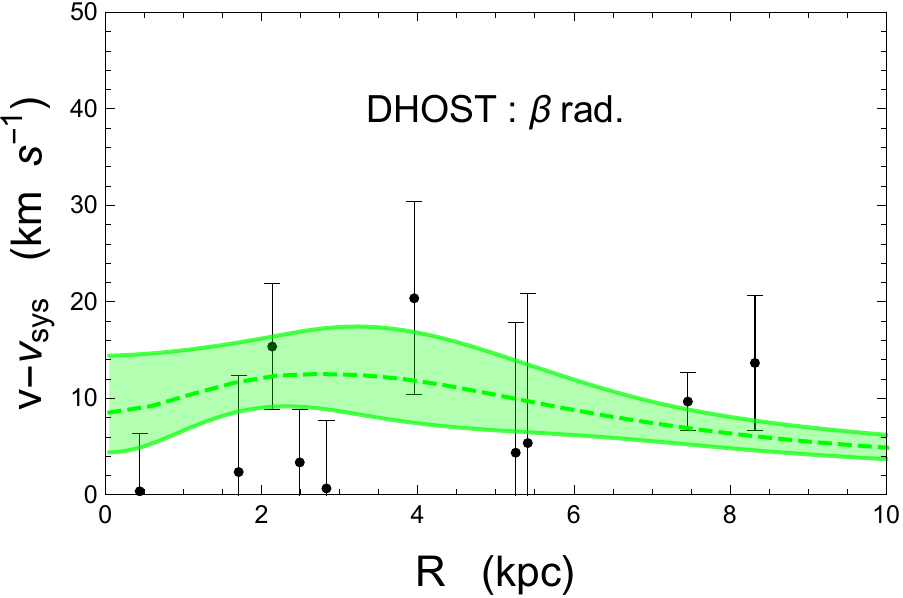}\
\caption{Velocity offset profiles of NGC1052-DF2 with DHOST. Black dots and bars are observational data, $v_{i}-v_{sys}$, with uncertainties, $\sigma_{v_{i}}$. Colored dashed line and shaded regions are respectively the median and the $1\sigma$ confidence region of the $\sigma_{los}$ profile derived from Eq.~\eqref{eqn: finvlos_DHOST}. Top panels: DHOST as DE component only. Bottom panels: DHOST replacing both DE and DM.}\label{fig:plot_DHOST}
\end{figure*}

This is the main clue on which is based the claim of NGC1052-DF2 being a ``lacking dark matter'' galaxy and is described in \cite{Wasserman:2018scp}. We also show in our Table~\ref{tab:results} that such scenario is only in mild tension with the reference star-only case. Which is quite the same as to say that the absence of DM at such scales in NGC1052-DF2 is a more than highly statistically valid hypothesis. We would like to point out also that we have considered the possibility that the gNFW might somehow fail to describe properly the DM halo of the galaxy (if any). We have thus relied also on a different DM model, the Einasto profile \cite{1965TrAlm...5...87E},
\be
\label{eqn: Einasto}
\rho_{Ein}(r) = \rho_s\exp\left\{-\frac{2}{\gamma}\left[\left(\frac{r}{r_s}\right)^\gamma - 1\right]\right\}\, ,
\ee
which is another three-parameters model which successfully applies to galactic scales, but absolutely no qualitative difference has come out.

We now turn our attention to the DHOST model, as our main goal is to test, and possibly show, its viability in describing the kinematical data of NGC1052-DF2. We first consider the scenario in which the DHOST theory plays the role of DE, whose effects might be felt at galactic scale by the breaking of the screening mechanism. In this case, we still need to include in the matter budget a DM component, which is always parametrized by a gNFW model. 

As it can be seen from Table \ref{tab:results} and from the top panels of Fig.~\ref{fig:plot_DHOST}, the presence of the DHOST model does not really change the results we get in the GR case, which are actually statistically equivalent. Indeed, the value of the DHOST parameter $\Xi_1$ is consistent with zero: as it might be expected, DE does not play a relevant role on galactic scales. On the other hand, we see that the addition of the DHOST model slightly raises both the Bayes factor and the suspiciousness with respect ot the case GR+DM: even if they remain mildly disfavoured, the tension is somehow alleviated, and this effect is most noticeable when the SHMR prior is applied.

The same conclusions can be driven in the case in which the DHOST plays both the role of DE at cosmological scales and entirely mimic DM at galactic ones. Results are shown both in Table \ref{tab:results} and in the bottom panels of Fig.~\ref{fig:plot_DHOST}. The most interesting thing to note is that, although the parameters are more or less statistically equivalent to the GR star-only case, and although the characteristic DHOST parameter has a non-zero median value but is consistent with the GR limit at $1\sigma$ confidence level, we have a raise in both the Bayes factor and in the suspiciousness. In the case of constant anisotropy the Bayes factor is even slightly positive, although by a negligible amount, for which we cannot really conclude that it should be preferred with respect to the reference GR-based case. The suspiciousness also becomes positive (only case among all the ones we have considered), meaning that it is fully consistent and not in tension with our reference model. 

That is the main conclusion and goal of our work: we have shown that even in a galaxy with a very low content of DM, or even lacking DM at all, DHOST theories cannot be discarded, but can be as much successful as GR in explaining observational data. Thus, NGC1052-DF2 has not dealt any deathblow to DHOST theories, which can be still be investigated as reliable ETGs candidates.

One further interesting and important point to address, is that in the case in which the DHOST plays also the role of DM, the chains, given only a limited number of physically reasonable priors, automatically set a sharp upper limit on the possible values of the characteristic DHOST parameters, namely $\Xi_{1}\lesssim0.5$ at $2\sigma$ confidence level. 

\begin{figure*}
\centering
\includegraphics[width=17cm]{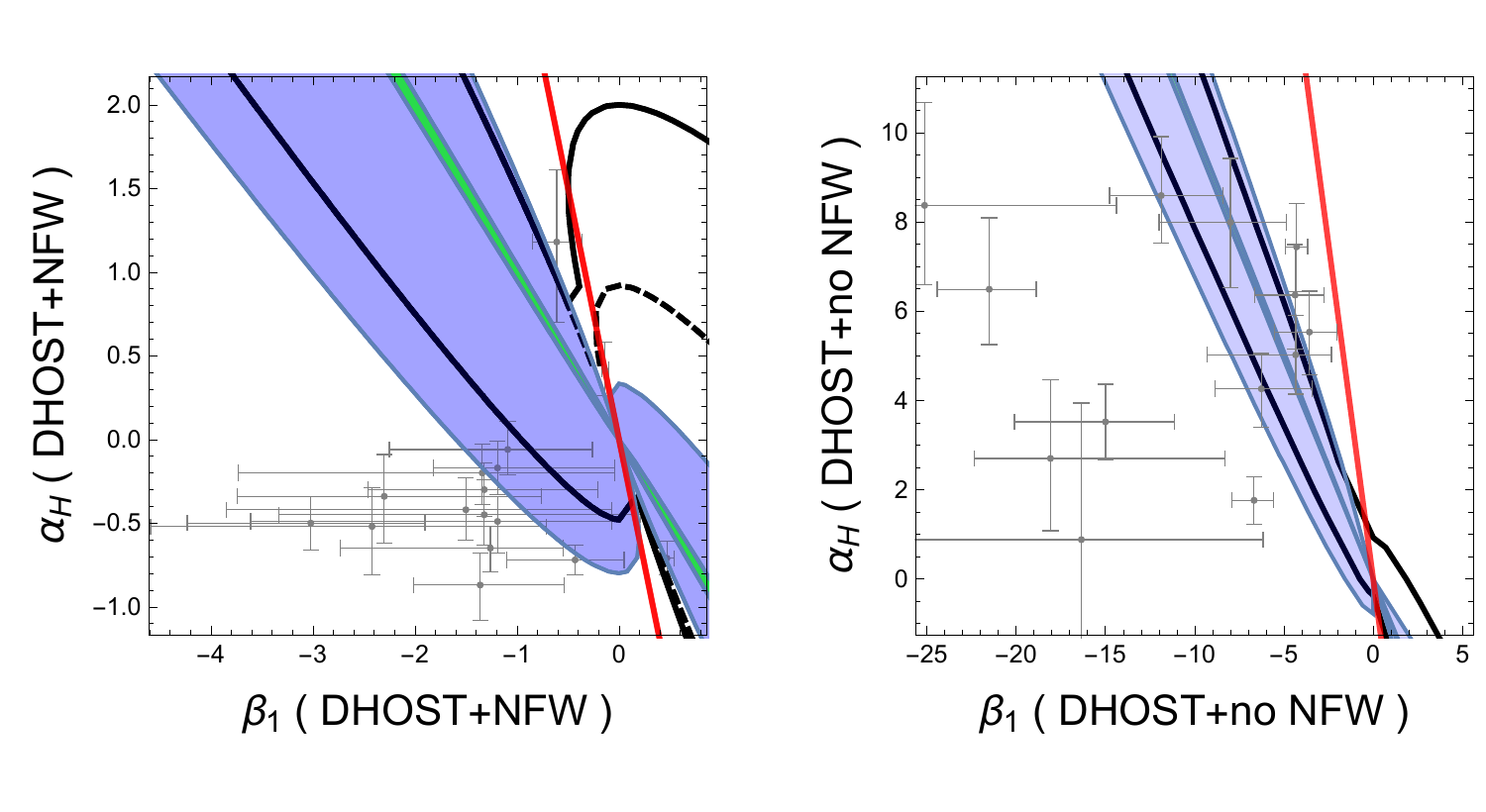}
\caption{Comparison of ETF parameter constraints from our DHOST analysis with the results from \cite{Dima:2017pwp}, \cite{Saltas:2019ius} and \cite{10.1093/mnras/stac180}. Left panel: results under the assumption that DHOST mimicks dark energy. Dashed black lines are $1\sigma$ constraints from this work for the case with SHMR prior; solid black lines are $1\sigma$ constraints from this work for the case without the SHMR prior. Right panel: results when DHOST is assumed to play the role of both dark energy and dark matter. Solid black lines are $1\sigma$ constraints from this work. In all cases we assume a constant stellar anisotropy profile. In both panels blue regions are derived from stellar physics considerations \cite{Dima:2017pwp}; the red region is derived from $2\sigma$ limits on $\gamma_0$ from the Hulse-Taylor pulsar; helioseismology $2\sigma$ constraints \cite{Saltas:2019ius}
are shown as green regions. Single constraints from CLASH clusters as obtained by \cite{10.1093/mnras/stac180} are shown as grey points/crosses.} \label{fig:EFT}
\end{figure*}

Finally, in Fig.~\ref{fig:EFT} we compare our constraints on $\Xi_1$, as function of the ETG parameters $\alpha_H$ and $\beta_1$, with others which are in literature. In blue, we have the constraints obtained from stellar physics arguments as conditions for dynamical equilibrium and for a minimal mass of red dwarf stars; in red, the $2\sigma$ limits on $\gamma_0$ from the Hulse-Taylor pulsar \cite{Dima:2017pwp}; in green, the constraints provided by helioseismology arguments \cite{Saltas:2019ius}; grey points with error bars represent the results we got in \cite{10.1093/mnras/stac180} from the analysis of the CLASH galaxy clusters. The new constraints on $\alpha_H$ and $\beta_1$ which can be derived from our estimations for $\Xi_1$ in this work, are represented by dashed black lines for the case with the SHMR prior, and solid black ones for the case without the SHMR prior, both with constant anisotropy profile. We can conclude that our new constraints are perfectly consistent with literature.

\section{Conclusions}
\label{sec: conclusions}

In this work we proceeded with our personal analysis of the DHOST model introduced in \cite{Crisostomi:2016czh, PhysRevD.97.021301, PhysRevD.97.101302}. This model has one interesting feature: a breaking of the corresponding screening mechanism which might help to unify DE and DM under one single theoretical scenario. In \cite{10.1093/mnras/stac180} we started our journey at clusters of galaxies' scales; here we tried to shed light on the dynamics of the Ultra-diffuse galaxy NGC1052-DF2 which has been claimed to be a ``lacking DM galaxy'' \cite{vanDokkum:2018vup} and, as such, might be a potential big problem for ETGs.

We tested the DHOST model described by Eq.~\eqref{eqn: model} in two different scenarios: one more conservative, in which the selected ETG only plays the role of DE at cosmological scales, but could alleviate the weight of DM by its broken screening mechanism; and one more ambitious, in which the DHOST model substitute entirely DM, thus playing the role of an ``effective'' mass. In the latter case we thus assume that the mass of NGC1052-DF2 is composed only by the baryonic (stellar) component.

We infer the mass of NGC1052-DF2 using the Jeans equation, Eq.~\eqref{eqn: Jeans}, and we model the galaxy's total mass as the sum of DM (when DM is assumed) and a stellar contribution. We describe DM using a generalization of the classical Navarro-Frenk-White profile. In addition, we consider two different models for the anisotropy parameter $\beta$: a case in which it is constant, and one with a radial profile described by Eq.~\eqref{eqn: anis}.

In agreement with results from \cite{Wasserman:2018scp}, we find that when GR is assumed, the best match with the data is obtained when no DM is included at all. While in \cite{Wasserman:2018scp} the authors always include a DM component, and conclude that it should be present in a very low amount, in this work we also explicitly consider the case with no DM, thus having a purely baryonic galaxy. We can infer that, at least at the scales tested by the observations, the hypothesis of a total DM absence might be considered a totally satisfying option, from the statistical point of view. Indeed, the inclusion of DM only makes worse all the Bayesian indexes we have considered (Bayes factor and suspiciousness). Such results are also quite independent on the priors which can be adopted.

When the DHOST model is assumed to act only as DE, we have a substantial equivalence between this scenario and the corresponding GR cases, which is somehow expected because if we have a DM component, the large scale effects of DE might be expected to be negligible at galactic scales. But we need to notice that all the Bayesian indicators are improved with respect to the GR+DM cases. Even so, the reference case of GR with only stars is still the most favoured, statistically speaking.

Finally, we assume the DHOST model as fully mimicking DM: when considering a constant anisotropy profile, we even get both a positive Bayes Factor and a positive suspiciousness. Although they are only slightly greater than zero (but different from it at least at $1\sigma$), they clearly point to the fact that NGC1052-DF2 can be quite satisfactory described by our DHOST model, as much successful as by GR.

Of course, any further conclusion cannot be definitely driven here, because the sample of analyzed objects is too small. But we are planning to extend it (and we have already started to work on that), so to include more UDGs with resolved kinematics, i.e. with data accurate enough to perform a kinematical analysis, such as those described in \cite{2019ApJ...874L...5V,PinaMancera:2021wpc,Shi:2021tyg,Kong:2022oyk}. Even more interestingly, UDGs are the perfect test arena for ETGs because they exhibit a wide range of behaviours, such that in the same family we can enlist also objects which seem to be highly DM dominated \cite{vanDokkum:2019fdc,Wasserman:2019ttq}. Analysis of these case are left to forthcoming papers.

{\renewcommand{\tabcolsep}{1.mm}
{\renewcommand{\arraystretch}{1.5}
\begin{table*}
\begin{minipage}{\textwidth}
\centering
\caption{Results from the statistical analysis of NGC1052-DF2. For each parameter we provide the median and the $1\sigma$ constraints; unconstrained parameters are in italic font.
The parameters are, from left to right: distance $D$; mass-to-light ratio $\Upsilon$; systemic velocity $v_{sys}$; anisotropy function parameters, depending on the model assumed, constant $(\beta_{c})$ or radial from \cite{Zhang:2015pca} $(\beta_{0},\beta_{infty},r_{a})$; gNFW concentration $c_{200}$, mass $M_{200}$, and inner log-slope $\gamma$; DHOST characteristic scaling $\Xi_1$; Bayes factor $\mathcal{B}_{i}^{j}$; its logarithm; and the suspiciousness $\log \mathcal{S}_{i}^{j}$.}\label{tab:results}
\resizebox*{\textwidth}{!}{
\begin{tabular}{c|ccc|cccc|cccc|ccc}
\hline
\hline
 & \multicolumn{11}{c}{GR}   \\
\hline
 & $D$ & $\Upsilon_{\ast}$ & $v_{sys}$ & $\beta_c$ & $\beta_0$ & $\beta_{\infty}$ & $r_a$ & $c_{200}$ & $\log M_{200}$ & $\gamma$ & $\Xi_1$ & $\mathcal{B}^{i}_{j}$ & $\log \mathcal{B}^{i}_{j}$ & $\log \mathcal{S}^{i}_{j}$ \\
 & Mpc & & km s$^{-1}$ &  &  &  & kpc &  & M$_{\odot}$ &  &  &  &  & \\ 
\hline
\multirow{2}{*}{Star only} & $22.09^{+1.23}_{-1.18}$ & $1.81^{+0.47}_{-0.46}$ & $1804.08^{+2.61}_{-2.55}$ & $-3.64^{+2.33}_{-3.20}$ & $-$ & $-$ & $-$ & $-$ & $-$ & $-$ & $-$ & $\mathit{1}$ & $\mathit{0}$ & $\mathit{0}$ \\
& $22.11^{+1.21}_{-1.18}$ & $1.62^{+0.47}_{-0.45}$ & $1804.28^{+2.98}_{-3.02}$ & $-$ & $-3.55^{+1.92}_{-2.79}$ & $-1.23^{+1.27}_{-2.18}$ & $\mathit{22.0}$ & $-$ & $-$ & $-$ & $-$ & $0.42^{+0.01}_{-0.01}$ & $-0.88^{+0.03}_{-0.04}$ & $-0.61^{+0.04}_{-0.03}$ \\
\hline
\multirow{2}{*}{SHMR+NFW} & $21.86^{+1.21}_{-1.19}$ & $1.59^{+0.50}_{-0.50}$ & $1801.93^{+3.85}_{-3.81}$ & $-1.92^{+1.81}_{-3.56}$ & $-$ & $-$ & $-$ & $8.16^{+3.40}_{-2.29}$ & $10.82^{+0.16}_{-0.18}$ & $<0.41$ & $-$ & $0.060^{+0.003}_{-0.002}$ & $-2.82^{+0.04}_{-0.03}$ & $-2.33^{+0.07}_{-0.05}$ \\
& $21.83^{+1.22}_{-1.21}$ & $1.56^{+0.51}_{-0.50}$ & $1802.27^{+4.01}_{-4.15}$ & $-$ & $-1.21^{+1.05}_{-2.23}$ & $-0.64^{+0.92}_{-2.11}$ & $\mathit{19.1}$ & $7.82^{+3.22}_{-2.23}$ & $10.81^{+0.16}_{-0.18}$ & $<0.37$ & $-$ & $0.028^{+0.001}_{-0.001}$ & $-3.58^{+0.04}_{-0.04}$ & $-2.85^{+0.06}_{-0.06}$ \\
\hline
\multirow{2}{*}{no SHMR+NFW} & $22.15^{+1.18}_{-1.20}$ & $1.78^{+0.47}_{-0.46}$ & $1804.01^{+2.72}_{-2.74}$ & $-3.34^{+2.19}_{-3.26}$ & $-$ & $-$ & $-$ & $24.11^{+12.15}_{-8.54}$ & $<6.78$ & $\mathit{0.02}$ & $-$ & $0.65^{+0.02}_{-0.02}$ & $-0.44^{+0.04}_{-0.03}$ & $-0.30^{+0.03}_{-0.03}$ \\
& $22.12^{+1.20}_{-1.22}$ & $1.64^{+0.49}_{-0.47}$ & $1804.22^{+3.11}_{-3.05}$ & $-$ & $-3.43^{+1.94}_{-2.84}$ & $-1.24^{+1.30}_{-2.27}$  & $\mathit{22.5}$ & $25.22^{+12.47}_{-8.39}$ & $<6.06$ & $\mathit{0.62}$ & $-$ & $0.279^{+0.007}_{-0.008}$ & $-1.28^{+0.03}_{-0.03}$ & $-0.92^{+0.05}_{-0.05}$ \\
\hline
\hline
 & \multicolumn{11}{c}{DHOST (as dark energy)}   \\
\hline
 & $D$ & $\Upsilon_{\ast}$ & $v_{sys}$ & $\beta_c$ & $\beta_0$ & $\beta_{\infty}$ & $r_a$ & $c_{200}$ & $\log M_{200}$ & $\gamma$ & $\Xi_1$ & $\mathcal{B}^{i}_{j}$ & $\log \mathcal{B}^{i}_{j}$ & $\log \mathcal{S}^{i}_{j}$ \\
 & Mpc & & km s$^{-1}$ &  &  &  & kpc &  & M$_{\odot}$ &  &  &  &  &  \\ 
\hline
SHMR+NFW & $21.91^{+1.19}_{-1.26}$ & $1.61^{+0.50}_{-0.53}$ & $1802.07^{+3.86}_{-3.71}$ & $-1.45^{+1.59}_{-3.28}$ & $-$ & $-$ & $-$ & $8.76^{+3.63}_{-2.50}$ & $10.82^{+0.16}_{-0.18}$ & $<0.56$ & $-0.25^{+0.49}_{-0.21}$ & $0.26^{+0.01}_{-0.01}$ & $-1.35^{+0.04}_{-0.04}$ & $-0.41^{+0.07}_{-0.07}$ \\
\hline
no SHMR+NFW & $22.07^{+1.20}_{-1.17}$ & $1.70^{+0.48}_{-0.49}$ & $1804.20^{+2.83}_{-2.95}$ & $-2.88^{+2.26}_{-3.46}$ & $-$ & $-$ & $-$ & $22.29^{+11.49}_{-7.78}$ & $<7.87$ & $\mathit{0.097}$ & $-0.29^{+0.53}_{-0.71}$ & $0.63^{+0.02}_{-0.02}$ & $-0.46^{+0.03}_{-0.03}$ &  $-0.25^{+0.02}_{-0.03}$ \\
\hline
\hline
 & \multicolumn{11}{c}{DHOST (as dark matter)}   \\
\hline
 & $D$ & $\Upsilon_{\ast}$ & $v_{sys}$ & $\beta_c$ & $\beta_0$ & $\beta_{\infty}$ & $r_a$ & $c_{200}$ & $\log M_{200}$ & $\gamma$ & $\Xi_1$ & $\mathcal{B}^{i}_{j}$ & $\log \mathcal{B}^{i}_{j}$ & $\log \mathcal{S}^{i}_{j}$ \\
 & Mpc & & km s$^{-1}$ &  &  &  & kpc &  & M$_{\odot}$ &  &  &  &  &  \\
\hline
\multirow{2}{*}{Star only} & $22.14^{+1.21}_{-1.20}$ & $1.81^{+0.47}_{-0.45}$ & $1804.10^{+2.68}_{-2.67}$ & $-3.25^{+2.32}_{-3.21}$ & $-$ & $-$ & $-$ & $-$ & $-$ & $-$ & $-0.15^{+0.34}_{-0.32}$ & $1.05^{+0.03}_{-0.03}$ & $0.04^{+0.03}_{-0.03}$ & $0.10^{+0.03}_{-0.03}$ \\
& $22.04^{+1.18}_{-1.18}$ & $1.61^{+0.51}_{-0.48}$ & $1804.35^{+3.03}_{-2.98}$ & $-$ & $-3.30^{+1.93}_{-2.87}$ & $-0.99^{+1.14}_{-2.32}$ & $\mathit{22.32}$ & $-$ & $-$ & $-$ & $-0.29^{+0.47}_{-0.49}$ & $0.43^{+0.02}_{-0.01}$ & $-0.84^{+0.04}_{-0.03}$ & $-0.52^{+0.04}_{-0.03}$ \\
\hline
\hline
\end{tabular}}
\end{minipage}
\end{table*}}}

\end{document}